# Metrics based Workload Analysis Technique for IaaS Cloud

Sukhpal Singh
Computer Science & Engineering Department
Thapar University, Patiala, Punjab, India
ssgill@thapar.edu

Inderveer Chana
Computer Science & Engineering Department
Thapar University, Patiala, Punjab, India
inderveer@thapar.edu

*Abstract*— **The Dynamic Scalability of resources, a problem in Infrastructure as a Service (IaaS) has been the hotspot for research and industry communities. The heterogeneous and dynamic nature of the Cloud workloads depends on the Quality of Service (QoS) allocation of appropriate workloads to appropriate resources. A workload is an abstraction of work that instance or set of instances that are going to perform. Running a web service or being a Hadoop data node is valid workloads. The efficient management of dynamic nature resources can be done with the help of workloads. Until workload is considered a fundamental capability, the Cloud resources cannot be utilized in an efficient manner. In this paper, different workloads have been identified and categorized along with their characteristics and constraints. The metrics based on Quality of Service (QoS) requirements have been identified for each workload and have been analyzed for creating better application design.**

## I. INTRODUCTION

The term workload in the context of Cloud Computing is an abstraction of the use to which Cloud consumers put their Virtual Machines (VMs) on the Cloud environment [5]. For example, a desktop workload might be supporting a number of users logging on to interactive desktop sessions. A SAP (System Application and Products) workload might be a system of VMs working together to support an enterprise's SAP system [11]. It is very important to schedule workloads and it directly affects the entire resource utilization. The issues of workload scheduling have been drawing much attention from both scientific and industrial communities. The goal of workload scheduling is to optimize time and cost and improve resource utilization by organizing and optimizing the scheduling process. Cloud Computing promises dynamic scalability, flexibility and cost-effectiveness to fulfil evolving computing desires [1] [2]. To realize these promises, Cloud providers need to be able to quickly plan and provision computing resources, so that the capacity of the supporting infrastructure can closely match the needs of new Cloud workloads [3].

Cloud workloads require categorization, so that server workload is analyzed at the group level, rather than at the individual server level [5] [6], this helps to achieve a deeper understanding of workload characteristics and greater prediction accuracy [7]. Treating server performance data samples as multiple time series, used to identify server groups in which certain workload patterns appear in a group [8] [9].

Executing too many workloads on a single resource will cause workloads to interfere with each other and result in degraded and unpredictable performance which, in turn, discourages the users. Thus, the providers may evict existing resources or reject resource requests to maintain service quality, but it could make the environment even more unpredictable. On the other hand, users want their workloads done at minimal expense or, in other words, they seek to maximize their cost performance (or minimize workload completion time). To address this problem, new solution should be developed. To successfully schedule workloads, initially we need to understand the Cloud workload (e.g. transactional database, file server, web server, application server and batch data processing) [33]. Based on this, user can design their applications which can lead to maximization of the scaling advantage [9]. With the help of this, not only dynamic infrastructure scaling can be achieved but it will minimize the response time of elastic demand and maximize the throughput of requests. With the extended use of Cloud technologies, applications that are envisioned to be part of their workload may have more complicated workloads rather than traditional data center ones [10].

The paper is structured as follows: Related work has been presented in Section 2. In Section 3, Workload Identification and Analysis has been presented. Case Study has been presented in Section 4. The Conclusion and the Future Work have been presented in Section 5.

## II. RELATED WORK

Workload is a major concern to achieve high performance on Clouds. A Walfredo Cirne et al. [26] have established hypothetical models to create illustrative workload traces. Arlitt et al. [27] have analyzed the workload classification on Web servers. Cherkasova et al. [28] have conducted an analysis on broadcasting servers. Gmach et al. [29] have considered the workload for data centre applications. Bobroff et al. [30] have used regression models to predict workload deviations, in order to dynamically place VMs. Verma et al. [31] have suggested consolidating servers using association or peak cluster based assignment. Jerry Rolia [32] have proposed a trace-based workload predicting technique for capacity management, here feedback controller can be used to allocate resources based on the current system status and the time-varying workload in order to meet SLA (Service Level Agreement) [1] [10]. For example, if SLA is a function of round trip time and resources are rented based on CPU time, then feedback controller can be built to achieve minimal-cost rental (e.g., from an IaaS provider) of CPU resource while maintaining a sufficiently low round trip time level under the time-varying workload [11].

The workload characteristics of each application are gathered in order to be used as a comparison in the future for identifying unknown applications that are going to be inserted in the infrastructure and as predictors in a linear regression model. In [8] [10], an investigation is carried out on the possibility to enhance the temporal isolation among VMs concurrently running on the same core, by using the IRMOS (Interactive Realtime Multimedia Applications in Service Oriented Infrastructures) real-time scheduler [12], focusing on compute intensive and network intensive workloads. A framework recommended by Nezih et al. [13] to produce and register test workloads for assessing the virtualization performance of the public Cloud over internet. A different observation explained in [11] is that a new Cloud standard should not need a static configuration of software and hardware components because dynamic provision and de-allocation of resources as well as the pay-as-you-go model are the intrinsic types of these facilities. The workload features of Intel's vConsolidate and VMware vMmark in [14] are the offered benchmarks for virtualization. The portable workloads of vCon standard are a database, a web server and a mail server. Every workload should run in its own VM. In [5], the main focus is the scalability of VMs with mixed workloads and the effect on their performance when consolidated. None of these works have used metric based approach. To this end, it calls for mechanism to characterize and predict server workload continued. Different from these existing works, our study aims to identify frequent and repeatable workload patterns [4].

III. WORKLOAD IDENTIFICATION AND ANALYSIS

A. *Workload Identifiaction*

Workloads need to be identified for the analysis and classification. The following Cloud workloads have been identified from literature along with their quality attributes [1] [3] [5] [6] [10] [11] [18] [19].

- Websites: Freely available websites for social networking, informational web sites large number of users. The Quality Attributes (QAs) for this workload are large amounts of reliable storage, high network bandwidth, performance and high availability.
- Technological Computing: It includes bioinformatics, atmospheric modeling, and other numerical computation. The QAs for this workload is computing capacity.
- Endeavour Software: It includes email servers, SAP, enterprise content management. The QAs for this workload are security, high availability, customer confidence level and correctness.
- Performance Testing: It includes simulation of large workloads to test the performance characteristics of software under development. The QAs for this workload is computing capacity and performance.
- Online Transaction Processing: It includes online insurance policies and online banking. The QAs for this workload are security, high availability, internet accessibility and usability.

- E-Commerce (E-Com): It includes super marketing. The QAs for this workload are variable computing load and customizability.
- Central Financial Services: It includes banking and insurance systems. The QAs for this workload are security, high availability, changeability and Integrity.
- Storage and Backup Services: It includes general data storage and backup. The QAs for this workload are reliability and persistence.
- Productivity Applications: It includes users signing up for mails, word editors. The QAs for this workload are network bandwidth, latency, data backup and security.
- Software/Project Development and Testing: It includes software development of web applications with Rational Software Architect, Microsoft Visual Studio etc. The QAs for this workload are user self-service rate, flexibility and testing time.
- Graphics Oriented Applications: It includes animation and visualization software applications. The QAs for this workload are network bandwidth and latency, data backup and visibility.
- Critical Internet Applications: It includes web applications including huge amount of scripting languages. The QAs for this workload are high availability, serviceability and usability.
- Mobile Computing Services: It includes Servers to support rich mobile applications. The QAs are portability, high availability and reliability.

B. *Workload Analysis*

A distinct workload (or a whole application) used by a set of consumers and a smaller facility may be used in dissimilar environments. The different applications have different set of requirements and characteristics. Some Clouds are natural fits for certain classes of workloads (i.e. Web Applications) whereas for another type of workloads (i.e. Batch), other Cloud services (Amazon Web Service) are more necessary [15]. The aim of workload analysis is to look at different aspects or characteristics of an enterprise application to determine the feasibility of moving or porting the application in the Cloud. This analysis also provides input to execution method, Cloud service choice and a preliminary business worth valuation [16] [17].

*1) Workload:* A workload can be minor or whole application. Industrial communities have to energetically handle workloads so to check how applications are running. The abstraction is a technique to retain the procedural information away from the consumer [5]. The outcome of this abstraction is a sort of service that makes it easier to have a distinct function with a defined determination [11]. The services live in a container with an Application Programming Interface (API) so it can be simply relocated from one place to another. Workloads are an important distinctive distinguishing the requirements for Cloud Computing [6].

*2) Workload Constraints:* The following are some constraints with respect to workloads:

   *a)* A workload may be time bound ("run for 1 hours") or time unbound.

*b)* A workload may have a particular begin time or elastic begin time.

*c)* A workload may have hard stop time (For Example: must finish by a certain time in the future).

*d)* A workload can be interruptible or must run without interruptions.

*e)* A workload may have a certain lower limit of the resource that it needs.

*f)* A workload may have urgency associated with it.

*g)* A workload may have budget associated with it.

*3) Workload Cost:* It includes the hardware cost, software cost, application maintenance and provision charges etc.

*4) Workload Characteristics: The following are* characteristics *for every Cloud workload:*

*a) Unstable Demand:* When a workload has an unchanging and sporadic demand, having dictated and well sized structure for that workload is possibly extra effective than reimbursing hourly charges for VMs in a public Cloud or building and using a private and automatic Cloud.

*b) Standard:* Usefulness in Cloud Computing is realized appreciations to virtualization and automation. Automation is budget effective if there is a lesser set of features (in SaaS services) or sections of software (in PaaS or IaaS services) present in the directory.

*c) Self-governing:* If a workload needs heavy communication with another system, relocations of that workload only to a public Cloud situation might affect performance undesirably because of concerns with latency and bandwidth between the data center and public Cloud situation. Though bandwidth can every time be increased, latency is harder to decrease lower than a lowest threshold unless 1 and 0 can travel faster than light.

*d) Not Critical:* The workloads with extraordinary challenging necessities (for example availability, response time, recovery time, recovery point and security) might not be organized to be presented in public Clouds so far. Service stages offered by public Clouds did not frequently fulfill the necessities of critical workloads.

*5) Workload Rate:* A workload may be used at the last days of the month or after every fixed time (periodically), workload occurrence is important for cost- benefit analysis.

*6) Workload as independent objects:* For independency, the following are features of these services:

- A workload has no dependencies. It's a distinct set of application logic that can be implemented individually of a particular application.
- The workload interface must be stable. Presently, the well-accepted interfaces are based on XML.
- A workload may have guidelines or strategies that apply in specific circumstances. There may be authorization and security strategies related with by a service for a specific function.

*7) Workload Categorization:* Practically four distinctive computing workloads based on different applications (consumer and wrapped) are identified and described.

*a)* CPU Oriented Workloads: These applications contain scientific calculation with important data munching, encryption and decryption, compression and decompression.

*b)* Memory Oriented Workloads: These applications contain in memory caching servers, in memory data servers.

*c)* Networking Oriented Workloads: These applications are web servers and network load balancers.

*d)* Storage Oriented Workloads: These applications include file serving and data mining.

*8) Workload Execution Modes:* The workload can be implemented in real time (i.e. online) as well as implemented at every time in a batch mode, summarized in Table I.

TABLE I. WORKLOADS BASED ON RESOURCE REQUIREMENT AND PROGRAMMING MODEL

| Criteria | Batch | Online |
|---|---|---|
| Resource Requirement | Require specific capacity in terms of storage and compute resources to finish the job in a timely manner. | The network bandwidth may be more critical. |
| Programming Model | Some batch jobs may be implemented over a framework like Hadoop. | For online workloads a PaaS may be the best option. |

*C. Workload Classification*

Different Cloud workloads have different features in terms of computing capacity, variability of load, network needs, back-up services, security needs, network bandwidth needs, and other QoS metrics [1] [3-8] [8]. Table II summarizes the common types of Cloud workloads. The new workloads made possible for Cloud are collaborative care, medical imaging, financial risk, energy management video encoding, multiplayer online gaming, tower planning, data analytics, graphical information system, agriculture Cloud etc.

TABLE II. DISTRIBUTION OF WORKLOADS

| Group | Workloads |
|---|---|
| Server Oriented | Websites, Technological computing, Endeavour software, Performance testing, Online transaction processing, E-Com, Central financial services, Storage and backup services. |
| Client Oriented | Production applications, Software/Project Development and testing, Graphics Oriented, Critical Internet applications. |
| Mobile Oriented | Mobile Computing services |

*D. Metrics for QoS Requirement*

Cloud workloads can be managed and executed efficiently with the help of metrics to measure the quality attributes for each workload. From the literature following metrics have been identified that can be applied to Cloud workloads [2] [20-29]. The abbreviations used in these metrics are: MTBF: Mean Time between Failures, MTTF: Mean Time to Failure, MTTR: Mean Time to Repair and MTTC: Mean Time to Change with respect to particular Cloud Service.

*A. Network Bandwidth*

The network bandwidth can be calculated as number of bits transferred/received in a particular workload in one second.

$$\text{Network Bandwidth} = \text{Bits/second (bps)}$$

### B. Integrity

$$\text{Integrity} = \sum [(1 - \text{Threat}) \times (1 - \text{Security})]$$

Where, the Threat is the probability of occurrence of an attack (of specific type) in given time and Security is the probability of repelling the specific attack in Cloud service.

### C. Usability

Work required knowledge, handling, preparing input, and interpreting output of a Cloud Service. It also includes on-line feedback and Service understand ability, Interface and aesthetic features of Service, Learnability: 1/(Time taken to learn the service) and Success ratio: (no of successful operations in a workload) / total operations available in the workload.

### D. Reliabiilty

$$\text{Reliability} = \text{MTBF} = \text{MTTF} + \text{MTTR}$$

### E. Availability

$$\text{Availability} = \text{MTTF} / \text{MTBF}$$

### F. Changeability

Changeability (MTTC) = ∑ (Time to analyze the change in workload + Time to modify the change in workload + Time to test the change in workload + time to distribute the change in workload) / (No. of change requests in workload).

### G. Latency

Latency = Time of input a Cloud workload – Time of output produced with respect to that Cloud workload.

### H. Cloud Customer Confidence Level

The Confidence and Fulfillment Matrix based on satisfaction level of Cloud Service as shown in Table III.

TABLE III.    CONFIDENCE AND FULFILLMENT MATRIX

| Fulfilment Level | Confidence (%) |
|---|---|
| Very Satisfied | 100 |
| Satisfied | 75 |
| Neutral | 50 |
| Dissatisfied | 25 |
| Completely Dissatisfied | 0 |

### I. Customizability

$$\frac{N_d}{N_d + N_s}$$

$N_d$: Number of dynamic changes in a Cloud service w.r.t. a workload. $N_s$: Number of static changes in a Cloud service w.r.t. a workload.

### J. Testing time

Testing Time = Time to prepare test environment + Time to execute Test Suite for a Cloud workload. (Test Suite is collection of test cases).

### K. Variable Computing Load

Variable Computing Load = Change in Load Balancing (Δ LB)

Δ LB = Actual load at time t / Expected load at time t

Δ LB ⩽ 1 for efficient Cloud service

### L. User Self Service Rate

It is defined as the ratio of the number of online inquiries regarding workload/resource into support by the total amount of unique visitors to web self-service site or Cloud service (CS) to make a request regarding different or same workload.

Self Service Rate = (100% − (# of Inquiries regarding workloads / # of Support Resource Visits)).

### M. Reliable Storage

- How and where do you store data of service?
- For how long do you store?
- How much do you store?
- Data mining: It should not be promising to make interpretations based on accessible design of CS.

### N. Database Backup

Database Backup for every CS = Giga Bytes (GB)

### O. Correctness

The degree to which the Cloud Service (CS) will be provided accurately to the Cloud customers.

Accuracy: (expected CS - | expected CS − observed CS|) /expected CS

Completeness: total existing CSs / total requested CSs

Defects/CS: Number of Defects reported/CS

### P. Service Visibility

Service Visibility: Degree of transparency for billing [33].

### Q. Serviceability

Serviceability = Service Uptime/ (Service Uptime + Service Downtime).

### R. Computing Capacity

Computing Capacity = Actual Usage time of the Resource/Expected Usage time of the Resource.

### S. Internet Accessibility (IA)

IA is defined as the ratio of the number of request time out to the total number of requests for a particular resource or service in the response of particular workload.

IA = Number of request time out/ Total number of requests or

IA= Percentage of Time Out Request

### T. Portability

Portability of service = Degree to which the service or CS is portable to other platforms.

Portability = (No. of compatible platforms)/ (total no. of platforms)

### U. Persistence

The number of time periods required for a given proportion of the total uncertainties in a given service to collect.

*V. Security*

Security [16] of CS can be measured by Security Metric Matrix as shown in Table IV.

TABLE IV. SECURITY METRIC MATRIX

| Security Measure Or Metric | Business Drivers | | | | | | |
|---|---|---|---|---|---|---|---|
| | CM | RM | RV | LR | PR | LS | II |
| The number of Fake alarms monitored by Corporate Security | X | X | X | | X | | |
| Security cost = % of total company revenue | X | X | | | | | X |
| Number of safety hazards proactively identified. | X | X | X | X | | | |
| % of dangerous data resources residing on systems. | X | X | X | | X | | |
| The number of ineffectual service responses to the issues identified by the Security as control weaknesses. | X | X | | X | | | X |
| Abbreviations: CM: Cost Management, RM: Risk Management, RV: ROI Value, LR: Legal Requirement, PR: Policy Requirement, LS: Life Safety and II: Internal Influence. | | | | | | | |

*W. Performance*

The workloads include for benchmark should be based upon a collection of the relevant datacentre workloads, and the performance metrics has been summarized in Table V.

TABLE V. PERFORMANCE METRICS BASED ON DIFFERENT SERVERS

| Workload | Metric |
|---|---|
| Mail server | No. of Actions/minute |
| Java server | No. of New orders/second |
| Web server | No. of Accesses/second |
| Database server | No. of Commits/second |
| File server | MB/second |

*X. Flexibility*

Flexibility: Service flexibility evaluation as follows:

**Flexible Point** $FXP_i$**:** A point or a location in service which can cause flexible changes to occur, upon which the external force $F_e$ may apply. $F_e$ causes software to change through the flexible point. Small external force $F_e$ at a $FXP$ may cause a large scale of changes in service. When $F_e=0$, it indicates that the service changes are completely driven by internal force $F_i$.

**Flexible Force** $f_i$: minimum external force $F_e$ applied to $FXP_i$ that may cause service to change. $f_i$ indicates the easiness or difficulty to make service change. The larger fi is, the harder the service makes changes through $FXP_i$.

**Flexible Distance** $S_i$ : maximum range or size of the service change caused by $f_i$ through flexible point $i$.

$$K_i: K_i = \frac{S_i}{1+f_i}$$

**Flexible Degree**, a measure for service flexibility at $FXP_i$.

**Flexible Capacity** $C$:

$$C = \sum_{i=1}^{N} K_i,$$

A measure of entire service flexibility, based on definitions above, a provider can utilize the flexibility at $i$ only if $F_e \geq F_i$.

IV. CASE STUDY

*Case 1:* Cloud workload (E-Commerce) can be scattered properly with the help of metric (Variable Computing Load). In the first scenario, a set of Cloud service customers concurrently uses Cloud Service C1, which is presented by Physical Server C1. Another physical server is existing but is not being used as shown in Figure 1.

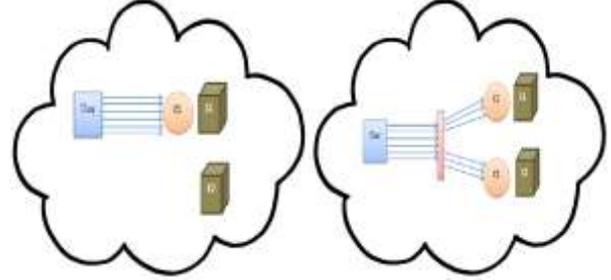

Figure 1. Unbalanced and balanced Workload

*Solution:* After measuring the Variable Computing Load (Metric identified in Section III), load is distributed on two servers. As an outcome, Physical Server C1 is over-utilized. In the second scenario, a redundant replica of Cloud Service C2 is executed on Physical Server C2. The load balancer diverts the Cloud service customer demands and points them to both Physical Server C1 and C2 to guarantee even scattering of the workload. With the help of this metric the workload (E-Com) will be scheduled and executed efficiently on the given resources; hence better resource utilization will be achieved.

*Case 2:* Synchronization in arriving of Cloud Workloads
*Solution:* After measuring the Latency (Metric) of every Graphic Oriented workload, the workload can be executed properly and resources will be scheduled efficiently hence it will improve execution time and cost. The structure of scheduled workload is shown in Figure 2 and the structure of non-scheduled workload is shown in Figure 3.

| WorkloadId | ProcessId | Begin Time | End Time | ResourceId |
|---|---|---|---|---|

Figure 2. Structure of Scheduled Workloads

| WorkloadId | ProcessId | Execution time | Resource List |
|---|---|---|---|

Figure 3. Structure of Non Scheduled Workloads

WorkloadId: It represents the identity of workloads. ProcessId: It represents the process identity. Execution Time: It represents the processing time of operation. Resource List: It represents the resource which can be used to execute this operation. Begin time: It represents the start time of operation. End time: It represents the end time of a particular operation. Resource id: It represents the resource which is used to execute the operation.

## V. CONCLUSION AND FUTURE SCOPE

In this paper, various Cloud workloads have been identified and classified along with their characteristics and restrictions. The metrics based on Quality of Service (QoS) requirements have been identified for every Cloud workload that can efficiently improve resource utilization. In the future work, the resource scheduling algorithm based on clustering of Cloud workloads can be incorporated to achieve better resource utilization along with time and cost optimization.

## REFERENCES

[1] P. Xiong, Zhikui Wang, Simon Malkowski, Qingyang Wang, Deepal Jayasinghe, and Calton Pu. "Economical and robust provisioning of n-tier Cloud workloads: A multi-level control approach." In Distributed Computing Systems (ICDCS), 2011 31st International Conference on, pp. 571-580. IEEE, 2011.

[2] M. Kunz, Andreas Schmietendorf, Reiner Dumke, and Cornelius Wille. "Towards a service-oriented measurement infrastructure." In Proc. of the 3rd Software Measurement European Forum (SMEF), pp. 197-207. 2006.

[3] K. Tsakalozos, Mema Roussopoulos, Vangelis Floros, and Alex Delis. "Nefeli: Hint-based execution of workloads in Clouds." In Distributed Computing Systems (ICDCS), 2010 IEEE 30th International Conference on, pp. 74-85. IEEE, 2010.

[4] V. D. Bossche, Ruben, Kurt Vanmechelen, and Jan Broeckhove. "Online cost-efficient scheduling of deadline-constrained workloads on hybrid Clouds." Future Generation Computer Systems (2012).

[5] O. Khalid, Ivo Maljevic, Richard Anthony, Miltos Petridis, Kevin Parrott, and Markus Schulz. "Dynamic scheduling of virtual machines running HPC workloads in scientific grids." In New Technologies, Mobility and Security (NTMS), 2009 3rd International Conference on, pp. 1-5. IEEE, 2009.

[6] Singh, Sukhpal, and Inderveer Chana. "Energy based Efficient Resource Scheduling: A Step Towards Green Computing." International Journal of Energy, Information & Communications 5, no. 2 (2014).

[7] G. Kousiouris, Tommaso Cucinotta, and Theodora Varvarigou. "The effects of scheduling, workload type and consolidation scenarios on virtual machine performance and their prediction through optimized artificial neural networks." Journal of Systems and Software 84, no. 8 (2011): 1270-1291.

[8] S. Mahambre, Purushottam Kulkarni, Umesh Bellur, Girish Chafle, and Deepak Deshpande. "Workload Characterization for Capacity Planning and Performance Management in IaaS Cloud." In Cloud Computing in Emerging Markets (CCEM), 2012 IEEE International Conference on, pp. 1-7. IEEE, 2012.

[9] S. Singh and Inderveer Chana. "Cloud Based Development Issues: A Methodical Analysis." International Journal of Cloud Computing and Services Science (IJ-CLOSER) 2, no. 1 (2012): 73-84.

[10] A. Nahir, Ariel Orda, and Danny Raz. "Workload factoring with the Cloud: A game-theoretic perspective." In INFOCOM, 2012 Proceedings IEEE, pp. 2566-2570. IEEE, 2012.

[11] A. Khan, Xifeng Yan, Shu Tao, and Nikos Anerousis. "Workload characterization and prediction in the Cloud: A multiple time series approach." In Network Operations and Management Symposium (NOMS), 2012 IEEE, pp. 1287-1294. IEEE, 2012.

[12] T. Cucinotta, Dhaval Giani, Dario Faggioli, and Fabio Checconi. "Providing performance guarantees to virtual machines using real-time scheduling." In Euro-Par 2010 Parallel Processing Workshops, pp. 657-664. Springer Berlin/Heidelberg, 2011.

[13] N. Yigitbasi, Alexandru Iosup, Dick Epema, and Simon Ostermann. "C-meter: A framework for performance analysis of computing Clouds." In Proceedings of the 2009 9th IEEE/ACM International Symposium on Cluster Computing and the Grid, pp. 472-477. IEEE Computer Society, 2009.

[14] M. A. El-Refaey, and Mohamed Abu Rizkaa. "Virtual systems workload characterization: An overview." In Enabling Technologies: Infrastructures for Collaborative Enterprises, 2009. WETICE'09. 18th IEEE International Workshops on, pp. 72-77. IEEE, 2009.

[15] M. Armbrust, Armando Fox, Rean Griffith, Anthony D. Joseph, Randy Katz, Andy Konwinski, Gunho Lee et al. "A view of Cloud Computing." Communications of the ACM 53, no. 4 (2010): 50-58.

[16] B. P. Rimal, Admela Jukan, Dimitrios Katsaros, and Yves Goeleven. "Architectural requirements for Cloud Computing systems: an enterprise Cloud approach." Journal of Grid Computing 9, no. 1 (2011): 3-26.

[17] L. Bass et al. (1997), Recommended Best Industrial Practice for Software Architecture Evaluation.

[18] http://www.1Cloudroad.com/Cloud-infrastructure-providers-for-2013.

[19] S. Singh, Inderveer Chana, "Consistency Verification and Quality Assurance (CVQA) Traceability Framework for SaaS", 3rd IEEE International Advance Computing Conference (IACC-2013), February 2013.

[20] L. Rosenberg, Ted Hammer, and Jack Shaw. "Software metrics and reliability." In Proceedings of the 9th International Symposium. 1998.

[21] L. H. Rosenberg, and Lawrence E. Hyatt. "Software quality metrics for object-oriented environments." Crosstalk Journal, April (1997).

[22] S. Zhao, Xiaoming Lu, Xianzhong Zhou, Tie Zhang, Jianqiang Xue, A Software Reliability Model For Web Services From the consumers' perspective, International Conference on Computer Science and Service System (CSSS), 2011, pp: 91-94.

[23] Singh, Sukhpal, and Inderveer Chana. "Formal Specification Language Based IaaS Cloud Workload Regression Analysis." arXiv preprint arXiv:1402.3034 (2014).

[24] L. J. Zhang, and Jia Zhang. "Architecture-driven variation analysis for designing Cloud applications." In Cloud Computing, 2009. CLOUD'09. IEEE International Conference on, pp. 125-134. IEEE, 2009.

[25] L. Tang, Jing Dong, Yajing Zhao, and Liang-Jie Zhang. "Enterprise Cloud service architecture." In Cloud Computing (CLOUD), 2010 IEEE 3rd International Conference on, pp. 27-34. IEEE, 2010.

[26] W. Crine and F. Berman, "A comprehensive model of the supercomputer workload," in Fourth Annual IEEE International Workshop on Workload Characterization. WWC-4, Austin, Texas, 2001.

[27] M. F. Arlitt and C. . L. Williamson, "Web server workload characterization: The search for invariants," in ACM SIGMETRICS international conference on Measurement and modeling of computer systems, SIGMETRICS '96, Philadelphia, PA, USA, 1996.

[28] Singh, Sukhpal, and Inderveer Chana. "Introducing Agility in Cloud Based Software Development through ASD." International Journal of u- and e-Service, Science and Technology 6, no. 191 (2013).

[29] D. Gmach, J. Rolia, L. Cherkasova and A. Kemper, "Workload analysis and demand prediction of enterprise data center applications," in IEEE 10th International Symposium on Workload Characterization, IISWC '07, Boston, MA, USA, 2007.

[30] N. Bobroff, A. Kochut and K. Beaty, "Dynamic placement of virtual machines for managing SLA violations," in IFIP/IEEE Integrated Network Management, Bombay, 2007.

[31] A. Verma , G. Dasgupta , T. Kumar and N. Prad, "Server workload analysis for power minimization using consolidation," in USENIX Annual technical conference, San Jose, CA, 2009.

[32] J. Rolia, . L. Cherkasova, . M. Arlitt and A. Andrzejak, "A capacity management service for resource pools," in 5th international workshop on Software and performance, Illes Balears, Spain, 2005.

[33] S. Singh. I. Chana, "Advance Billing and Metering Architecture for Infrastructure as a Service", International Journal of Cloud Computing and Services Science International Journal of Cloud Computing and Services Science (IJ-CLOSER) 2, no. 2 (2013): 123-133.